\documentclass[10pt,conference]{IEEEtran}
\usepackage{epsf,psfrag,amssymb,amsfonts,cite}

\newtheorem{example}{Example}

\newtheorem{definition}{Definition}

\def\psfancypar#1#2{\begingroup\def\par{\endgraf\endgroup\lineskiplimit=0pt}
               \setbox2=\hbox{\large\sc #2}
               \newdimen\tmpht \tmpht \ht2 \advance\tmpht by \baselineskip
               \font\hhuge=Times-Bold at \tmpht
               \setbox1=\hbox{{\hhuge #1}}
               \count7=\tmpht \count8=\ht1
               \divide\count8 by 1000 \divide\count7 by \count8 
               \tmpht=.001\tmpht\multiply\tmpht by \count7 
               \font\hhuge=Times-Bold at \tmpht
               \setbox1=\hbox{{\hhuge #1}}
               \noindent
                \hangindent1.05\wd1
               \hangafter=-2 {\hskip-\hangindent
               \lower1\ht1\hbox{\raise1.0\ht2\copy1}%
                \kern-0\wd1}\copy2\lineskiplimit=-1000pt}

\newcommand{\beq}{\begin{equation}}
\newcommand{\eeq}{\end{equation}}
\newcommand{\bqa}{\begin{eqnarray}}
\newcommand{\eqa}{\end{eqnarray}}
\newcommand{\bqn}{\begin{eqnarray*}}
\newcommand{\eqn}{\end{eqnarray*}}
\newcommand{\nn}{\nonumber}

\newcommand{\be}{\begin{enumerate}}
\newcommand{\ee}{\end{enumerate}}
\newcommand{\bi}{\begin{itemize}}
\newcommand{\ei}{\end{itemize}}
\newcommand{\bd}{\begin{description}}
\newcommand{\ed}{\end{description}}
\newcommand{\ba}{\begin{array}}
\newcommand{\ea}{\end{array}}
\newcommand{\bde}{\begin{definition}}
\newcommand{\ede}{\end{definition}}
\newcommand{\bex}{\begin{example}}
\newcommand{\eex}{\end{example}}

\def\boxit#1{\vbox{\hrule\hbox{\vrule\kern3pt
        \vbox{\kern3pt#1\kern3pt}\kern3pt\vrule}\hrule}}

\def\reals{ { {\rm  I \kern-0.15em R }  } }
\def\complex{ {\,{{\rm C} \kern-0.50em \raise0.20ex {  |}}\, }}

\def\0bf{{\bf 0}}
\def\1bf{{\bf 1}}
\def\2bf{{\bf 2}}
\def\3bf{{\bf 3}}
\def\4bf{{\bf 4}}
\def\5bf{{\bf 5}}
\def\6bf{{\bf 6}}
\def\7bf{{\bf 7}}
\def\8bf{{\bf 8}}
\def\9bf{{\bf 9}}

\def\gbf{{\bf g}}
\def\hbf{{\bf h}}

\def\mbf{{\bf m}}
\def\nbf{{\bf n}}

\def\qbf{{\bf q}}

\def\sbf{{\bf s}}

\def\ubf{{\bf u}}
\def\vbf{{\bf v}}
\def\wbf{{\bf w}}
\def\xbf{{\bf x}}
\def\ybf{{\bf y}}

\def\xbf{{\bf x}}
\def\ybf{{\bf y}}

\def\Rbf{{\bf R}}

\def\Cmat{\mathcal{C}}
\def\Dmat{\mathcal{D}}

\def\Gmat{\mathcal{G}}

\def\Mmat{\mathcal{M}}

\def\Pmat{\mathcal{P}}
\def\Qmat{\mathcal{Q}}
\def\Rmat{\mathcal{R}}

\def\Xmat{\mathcal{X}}
\def\Ymat{\mathcal{Y}}

\def\Rxx{\Rbf_{\ssstyle X\kern-.1em X}}

\let\ssstyle=\scriptscriptstyle

\def\Kout{\setbox1=\hbox{\Huge\bf K}\hbox to
1.05\wd1{\hspace{.05\wd1}
}}
\def\Sout{\setbox1=\hbox{\Huge\bf S}\hbox to 1.05\wd1{\hspace{.05\wd1}
}}

\begin{document}

\title{\bf \LARGE An Achievable Rate Region for Interference Channels \\with Conferencing}
\author{\authorblockN{Yi Cao and Biao Chen}
\authorblockA{Department of EECS, Syracuse University, Syracuse, NY 13244\\
Email: ycao01@syr.edu, bichen@ecs.syr.edu}} \maketitle
\begin{abstract}
In this paper, we propose an achievable rate region for discrete
memoryless interference channels with conferencing at the
transmitter side. We employ superposition block Markov encoding,
combined with simultaneous superposition coding, dirty paper
coding, and random binning to obtain the achievable rate region.
We show that, under respective conditions, the proposed achievable
region reduces to Han and Kobayashi's achievable region for
interference channels, the capacity region for degraded relay
channels, and the capacity region for the Gaussian vector
broadcast channel. Numerical examples for the Gaussian case are
given.
\end{abstract}

\noindent {\em Index terms} --- interference channels, dirty paper
coding, superposition block Markov encoding, random binning.
 \maketitle

 \section{Introduction}
 The capacity region of an interference channel (IC), where the
information sources at the two transmitters are statistically
independent, has been a long standing problem. Carleial was the
first to use the superposition code idea \cite{Carleial:78IT} to
obtain an inner bound for IC. This inner bound was later improved
by Han and Kobayashi \cite{Han&Kobayashi:81IT} who gave an
achievable rate region that is the largest reported to this date.
Recently, a simplified description of the Han-Kobayashi (HK) rate
region for the general IC is derived by Chong-Motani-Garg in
\cite{CMG:06IT}.

A related and less well investigated problem is when the
information sources at the two transmitters are correlated, i.e.,
interference channel with common information (ICCI)
\cite{Tan:80IC,Kramer:05Asilomar}. In \cite{Tan:80IC}, an
achievable rate region, an outer bound, and a limiting expression
for the capacity region were obtained. Later, the capacity region
of this channel under strong interference was found in
\cite{Kramer:05Asilomar}. Recently, improved achievable regions
for general ICCI \cite{Cao:07WCNC,JXG:06IT} and three new outer
bounds for the capacity region of Gaussian ICCI
\cite{Cao:07globecom} were proposed. However, all those results
are based on the assumption that the common message is available
noncausally.

In this work, we investigate the problem of user cooperation in
interference channels for the causal case. Here, each user not
only transmits his own message to the intended receiver, but also
serves as a relay to help transmit part of the other user's
message. We apply the superposition block Markov encoding, which
was used previously for the relay channel
\cite{Cover&Elgamal:79IT} and for user cooperation in multiple
access channels \cite{Sendonaris-etal:03COM}. Our proposed
achievable rate region is a generalized form of the HK region for
IC \cite{Han&Kobayashi:81IT}, the capacity region of degraded
relay channels \cite{Cover&Elgamal:79IT}, and the capacity region
of the Gaussian vector broadcast channel (GVBC) \cite{WSS:06IT}.

This paper is organized as follows. In section II, we present the
channel model and review some existing results. In section III, we
propose an achievable region for general IC with transmitter
conferencing.
%We also establish that, under certain conditions,
%the proposed region coincides with the HK region for general IC
%\cite{Han&Kobayashi:81IT}, the capacity region of degraded relay
%channel\cite{Cover&Elgamal:79IT}, and the capacity region of GVBC
%\cite{WSS:06IT}, respectively.
In section IV, numerical examples are used to compare the proposed
region with the HK region and the capacity region of GVBC. We
conclude in section V.

\section{Preliminaries and Existing Results}
\subsection{Definitions}
A memoryless discrete IC with conferencing (ICC) is denoted by
($\Xmat_1,\Xmat_2,p,\Ymat_1,\Ymat_2,\tilde{\Ymat}_1,\tilde{\Ymat}_2$),
where $\Xmat_1,\Xmat_2$ are two finite alphabet sets for the
channel input, $\Ymat_1,\Ymat_2$ are two finite
\begin{figure}[htp]
\centerline{\leavevmode \epsfxsize=3.5in \epsfysize=1.5in
\epsfbox{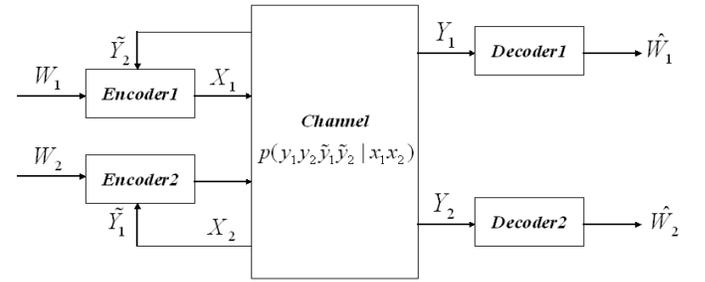}} \caption{\label{fig:model} Interference
channel with conferencing at the transmitter side.}
\end{figure}
alphabet sets for the channel output, $\tilde{\Ymat}_1,
\tilde{\Ymat}_2$ are two finite alphabet sets for the received
signals at the transmitters (which also serve as relays), and $p$
is the channel transition probability
$p(y_1,y_2,\tilde{y}_1,\tilde{y}_2|x_1,x_2)$. Here we assume the
channel is memoryless and encoders 1 and 2 are allowed to depend
only on their own messages and the past values of $\tilde{y}_2$
and $\tilde{y}_1$. Let $\Mmat_1=\{1,2,\cdot\cdot\cdot,M_1\}$ and
$\Mmat_2=\{1,2,\cdot\cdot\cdot,M_2\}$ be the message sets of
sender 1 and sender 2, respectively. Thus, for $w_1\in \Mmat_1$
and $w_2\in \Mmat_2$, the joint probability mass function of
$\Mmat_1\times\Mmat_2\times\Xmat_1^n\times\Xmat_2^n\times\Ymat_1^n\times\Ymat_2^n
\times\tilde{\Ymat}_1^n\times\tilde{\Ymat}_2^n$ is given by \beq
\begin{array}{ll}
p(w_1,w_2,\xbf_1,\xbf_2,\ybf_1,\ybf_2,\tilde{\ybf}_1,\tilde{\ybf}_2)\\
=p(w_1)p(w_2)\prod_{i=1}^np(x_{1i}|w_1,\tilde{y}_{21},\cdot\cdot\cdot,\tilde{y}_{2i-1})\\
\times
p(x_{2i}|w_2,\tilde{y}_{11},\cdot\cdot\cdot,\tilde{y}_{1i-1})p(y_{1i},y_{2i},\tilde{y}_{1i},\tilde{y}_{2i}|x_{1i},x_{2i})
\end{array}
\eeq Suppose $w_1\in \Mmat_1$  and $w_2\in \Mmat_2$ are sent by
transmitters 1 and 2 respectively, $g_1$ and $g_2$ are the
decoding functions at receivers 1 and 2; the average probabilities
of decoding error of this channel are defined as \bqa
P_{e,1}^{(n)}\equiv\frac{1}{M_1M_2}\sum_{w_1,w_2}Pr(g_1(Y_1)\neq
w_1|w_1,w_2 \hspace{0.2cm}\mbox{sent}) \\
P_{e,2}^{(n)}\equiv\frac{1}{M_1M_2}\sum_{w_1,w_2}Pr(g_2(Y_2)\neq
w_2|w_1,w_2 \hspace{0.2cm}\mbox{sent})\eqa The capacity region of
ICC is the closure of all rate pairs $(R_1,R_2)$ such that
$P_{e,1}^{(n)}\rightarrow 0, P_{e,2}^{(n)}\rightarrow 0$ as
codeword length $n\rightarrow\infty$, where $R_1=\frac{1}{n}\log
M_1$ and $R_2=\frac{1}{n}\log M_2$.

\subsection{Existing Results}
\emph{1)} Chong-Motani-Garg recently derived a simplified
description of the HK region for IC \cite{CMG:06IT}, as summarized
below.

\emph{Proposition 1}: Let $\Pmat_1^{*}$ be the set of probability
distributions $P_1^{*}(\cdot)$ that factor as \bqa
P_1^{*}(q,u_1,u_2,x_1,x_2)=p(q)p(x_1u_1|q)p(x_2u_2|q).\eqa For a
fixed $P_1^{*}\in \Pmat_1^{*}$, let $\Rmat_{HK}^{c}(P_1^{*})$ be
the set of $(R_1,R_2)$ satisfying (9)-(15) in Theorem 2 of
\cite{CMG:06IT}.
%\bqa
%R_1\!\!\!&\leq&\!\!\!I(X_1;Y_1|U_2Q)\\
%R_1\!\!\!&\leq&\!\!\!I(U_1;Y_2|X_2Q)+I(X_1;Y_1|U_1U_2Q)\\
%R_2\!\!\!&\leq&\!\!\!I(X_2;Y_2|U_1Q)\\
%R_2\!\!\!&\leq&\!\!\!I(U_2;Y_1|X_1Q)+I(X_2;Y_2|U_1U_2Q)\\
%R_1+R_2\!\!\!&\leq&\!\!\!I(X_1U_2;Y_1|Q)+I(X_2;Y_2|U_1U_2Q)\\
%R_1+R_2\!\!\!&\leq&\!\!\!I(X_1;Y_1|U_1U_2Q)+I(X_2U_1;Y_2|Q)\\
%R_1+R_2\!\!\!&\leq&\!\!\!I(X_1U_2;Y_1|U_1Q)+I(X_2U_1;Y_2|U_2Q)\\
%\nn 2R_1+R_2\!\!\!&\leq&\!\!\!I(X_1U_2;Y_1|Q)+I(X_1;Y_1|U_1U_2Q)\\
%\!\!\!&&\!\!\!+I(X_2U_1;Y_2|U_2Q)\\
%\nn R_1+2R_2\!\!\!&\leq&\!\!\!I(X_2;Y_2|U_1U_2Q)+I(X_2U_1;Y_2|Q)\\
%\!\!\!&&\!\!\!+I(X_1U_2;Y_1|U_1Q) \eqa
Then $\bigcup_{P_1^{*}\in\Pmat_1^{*}}\Rmat_{HK}^{c}(P_1^{*})$ is
equivalent to the HK region.

\emph{2)} The capacity of the degraded relay channel is given in
proposition 2 \cite{Cover&Elgamal:79IT}.

\emph{Proposition 2}: A relay channel consists of an input $x_1$,
a relay output $y_1$, a channel output $y$, and a relay sender
$x_2$ (whose transmission is allowed to depend on the past symbols
of $y_1$). If $y$ is a degraded form of $y_1$
\cite{Cover&Elgamal:79IT}, then\bqa
C=\max_{p(x_1,x_2)}\min\{I(X_1,X_2;Y),I(X_1;Y_1|X_2)\}.\eqa

\emph{3)} The capacity region of GVBC is computed using a
covariance matrix constraint on the inputs $X=(X_1,X_2)^T$ of the
form $E[XX^T]\leq S$. In order to mimic the individual power
constraints $P_1$ and $P_2$ on the two users for the vector case,
the input
covariance matrix $S$ is of the form $S=\left(%
\begin{array}{cc}
  P_1 & c \\
  c & P_2 \\
\end{array}%
\right)$, for some $-\sqrt{P_1P_2}\leq c\leq \sqrt{P_1P_2}$. Then,
the capacity region of GVBC is given below \cite{WSS:06IT}.

\emph{Proposition 3}: For each such S and all positive
semi-definite matrices $B$ and $D$, where $B+D\leq S$, both rate
pairs \beq
\begin{array}{ll}R_1\leq
\frac{1}{2}\log\left(\frac{|H_1BH_1^T+Q_1|}{|Q_1|}\right), R_2\leq
\frac{1}{2}\log\left(\frac{|H_2(B+D)H_2^T+Q_2|}{|H_2BH_2^T+Q_2|}\right)\end{array}\label{eq:Gaussian1}\eeq
and  \beq \begin{array}{ll}R_1\leq
\frac{1}{2}\log\left(\frac{|H_1(B+D)H_1^T+Q_1|}{|H_1DH_1^T+Q_1|}\right),
R_2\leq
\frac{1}{2}\log\left(\frac{|H_2DH_2^T+Q_2|}{|Q_2|}\right)\end{array}\label{eq:Gaussian2}\eeq
are achievable, where $H_1=(1,a_{21})$ and $H_2=(a_{12},1)$. The
convex hull of the union of these pairs over all possible $S, B$
and $D$ matrices is the capacity region of GVBC.

\section{Main Results}
We first give a brief outline of our encoding-decoding strategy.
We split each user's message into two parts: $M$ and $W$, where
$M$ is to be sent directly to the intended receiver, and $W$ is
the cooperative message to be sent to the receiver via the
cooperation of the other user (relay). Our cooperation strategy is
based on superposition block Markov encoding with the assumption
that $W$ can be perfectly decoded by the relay. The purpose of
introducing $M$ is to achieve a reasonable rate region (no less
than IC without conferencing) even when the conferencing channel
is poor. For the message $M$, we apply simultaneous superposition
coding \cite{Han&Kobayashi:81IT} and further split it into two
parts: private message $V$ and common message $U$.

For the cooperation in transmitting $W$, we jointly consider $B$
blocks, each of $n$ symbols. Each user transmits a sequence of
$B-1$ messages $w_1,\cdot\cdot\cdot,w_{B-1}$ in $B$ blocks, with
no new message in the last block. Note that as $B\rightarrow
\infty$, $(B-1)/B$ is arbitrarily close to $1$, hence the penalty
on rate is negligible. Now we take user 1 as an example to show
the whole process. Suppose there are $2^{nR_{13}}$ codewords of
$W_1$ for user 1 to transmit. We establish a random partition by
randomly throwing them into $2^{nR_{10}}$ cells. This partition is
made known to both transmitters and receivers. Suppose user 1
sends message $w_{1,b-1}$ at block $b-1$. At the end of block
$b-1$, following our assumption, user 2 can perfectly decode
$w_{1,b-1}$ and calculate the cell index $s_{1b}$ to which
$w_{1,b-1}$ belongs. At block $b$, both user 1 and 2 spend some
power transmitting $s_{1b}$. This provides the basis for
cooperatively resolving the remaining $Y_1$ uncertainty about
$w_{1,b-1}$. After decoding $s_{1b}$ at the end of block $b$,
receiver $Y_1$ intersects its ambiguity set $\Dmat(y_1(b-1))$
(i.e., the set of all codewords $w$ that are jointly typical with
$y_{1,b-1}$\cite{Cover&Elgamal:79IT}) with cell $s_{1b}$ and gets
the unique correct codeword $w_{1,b-1}$ with a probability close
to 1.

Since both users 1 and 2 can perfectly decode each other's
messages $W_1$, $W_2$ and then calculate the corresponding cell
indices $S_1, S_2$, we can employ dirty paper coding (DPC) to
transmit $V_1, U_1$ and $S_1$ treating $S_2$ as a known
interference at transmitter 1. Similarly at transmitter 2, we can
transmit $V_2, U_2$ and $S_2$ treating $S_1$ as a known
interference. Thus, introducing auxiliary random variables $M_1,
N_1, G_1, H_1$ and $M_2, N_2, G_2, H_2$ for DPC, we summarize the
achievable region for ICC in the theorem below.

\emph{Thoerem 1}: Let $Z_1=(Y_1, Y_2, \tilde{Y}_1, \tilde{Y}_2,
X_1, X_2, M_1, N_1, G_1,\\ H_1, V_2, U_2, W_2, S_2,Q)$ and let
$\Pmat_1^{*}$ be the set of distribution on $Z_1$ that can be
factored into the form \beq
\begin{array}{ll}
p(q)p(u_2|q)p(w_2|q)p(s_2|q)p(v_2|u_2s_2q)\\\times
p(n_1|s_2q)p(g_1|s_2q)p(h_1|s_2q)p(m_1|n_1h_1s_2q)\\\times
p(x_1|m_1g_1s_2q)p(x_2|v_2w_2h_1q)p(y_1y_2\tilde{y}_1\tilde{y}_2|x_1x_2)
\end{array}
\eeq Let $S(Z_1)$ be the set of $(R_1,R_2)$ such that
$R_1=R_{11}+R_{12}+R_{13}$ and $R_2=R_{22}+R_{21}+R_{23}$
satisfying: \bqa R_{11}\!\!\!&\leq&\!\!\! L_{11}-I(M_1;S_2|N_1H_1Q)\label{eq:1}\\
R_{12}\!\!\!&\leq&\!\!\! L_{12}-I(N_1;S_2|Q)\label{eq:2}\\
R_{13}\!\!\!&\leq&\!\!\! L_{13}-I(G_1;S_2|Q)\\
R_{10}\!\!\!&\leq&\!\!\! L_{10}-I(H_1;S_2|Q)\label{eq:4}\\
%\nn\\
L_{11}\!\!\!&\leq&\!\!\! I(Y_1N_1H_1U_2;M_1|Q)\label{eq:5}\\
%R_{21}\!\!\!&\leq&\!\!\! I(Y_1M_1N_1H_1;U_2|Q)\\
L_{11}+L_{12}\!\!\!&\leq&\!\!\! I(Y_1H_1U_2;M_1N_1|Q)\\
L_{11}+L_{10}\!\!\!&\leq&\!\!\! I(Y_1N_1U_2;M_1H_1|Q)\\
L_{11}+R_{21}\!\!\!&\leq&\!\!\! I(Y_1N_1H_1;M_1U_2|Q)\\
L_{11}+L_{12}+L_{10}\!\!\!&\leq&\!\!\! I(Y_1U_2;M_1N_1H_1|Q))\\
L_{11}+L_{12}+R_{21}\!\!\!&\leq&\!\!\! I(Y_1H_1;M_1N_1U_2|Q)\\
L_{11}+L_{10}+R_{21}\!\!\!&\leq&\!\!\! I(Y_1N_1;M_1H_1U_2|Q)\\
L_{11}+L_{12}+L_{10}+R_{21}\!\!\!&\leq&\!\!\! I(Y_1;M_1N_1H_1U_2|Q)\label{eq:13}\\
L_{13}\leq R_{10}\!\!\!&+&\!\!\!I(Y_1M_1N_1H_1U_2;G_1|Q)\label{eq:14}\\
L_{13}\!\!\!&\leq&\!\!\! I(\tilde{Y}_1H_1S_2;G_1|Q)\label{eq:15}\\
%\nn\\
R_{22}\!\!\!&\leq&\!\!\! I(Y_2U_2S_2N_1;V_2|Q)\label{eq:16}\\
%L_{12}\!\!\!&\leq&\!\!\! I(Y_2V_2U_2S_2;N_1|Q)\\
R_{22}+R_{21}\!\!\!&\leq&\!\!\! I(Y_2S_2N_1;V_2U_2|Q)\\
R_{22}+R_{20}\!\!\!&\leq&\!\!\! I(Y_2U_2N_1;V_2S_2|Q)\\
R_{22}+L_{12}\!\!\!&\leq&\!\!\! I(Y_2U_2S_2;V_2N_1|Q)\\
R_{22}+R_{21}+R_{20}\!\!\!&\leq&\!\!\! I(Y_2N_1;V_2U_2S_2|Q)\\
R_{22}+R_{21}+L_{12}\!\!\!&\leq&\!\!\! I(Y_2S_2;V_2U_2N_1|Q)\\
R_{22}+R_{20}+L_{12}\!\!\!&\leq&\!\!\! I(Y_2U_2;V_2S_2N_1|Q)\\
R_{22}+R_{21}+R_{20}+L_{12}\!\!\!&\leq&\!\!\!
I(Y_2;V_2U_2S_2N_1|Q)\\
R_{23}\leq R_{20}\!\!\!&+&\!\!\!I(Y_2V_2U_2S_2N_1;W_2|Q)\\
R_{23}\!\!\!&\leq&\!\!\!
I(\tilde{Y}_2H_1S_2;W_2|Q)\label{eq:26}\eqa Let
$\Rmat_1^{*}=\bigcup_{Z_1\in\Pmat_1^{*}}S(Z_1)$. Swap index 1 and
2 in all of the above statements and inequalities and we get
$\Rmat_2^*=\bigcup_{Z_2\in\Pmat_2^{*}}S(Z_2)$. Then the achievable
region $\Rmat^*=convhull(R_1^*\cup R_2^*)$ and the cardinality
$||Q||\leq 33$.
\\
\emph{Proof}: We only need to prove the achievability of $R_1^*$.

\emph{Codebook Generation}: Let
$\qbf=(q^{(1)},\cdot\cdot\cdot,q^{(n)})$ be a random sequence of
$\Qmat^{n}$ distributed according to $\prod_{t=1}^{n}p(q^{(t)})$.
Generate $2^{nR_{21}}$ i.i.d (independent and identically
distributed) codewords $\ubf_2(j_2)$ for common messages,
$2^{nR_{20}}$ i.i.d codewords $\sbf_2(l_2)$ for cell indices, and
$2^{nR_{23}}$ i.i.d codewords $\wbf_2(k_2)$ for cooperative
messages according to $\prod_{t=1}^np(u_2^{(t)}|q^{(t)})$,
$\prod_{t=1}^np(s_2^{(t)}|q^{(t)})$ and
$\prod_{t=1}^np(w_2^{(t)}|q^{(t)})$, respectively. For each pair
of $(\ubf_2(j_2),\sbf_2(l_2))$, generate $2^{nR_{22}}$ i.i.d
codewords $\vbf_2(i_2,j_2,l_2)$ for private messages according to
$\prod_{t=1}^np(v_2^{(t)}|u_2^{(t)}s_2^{(t)}q^{(t)})$\footnote{Note
that the codebook generation for the direct transmission part
follows that of \cite{CMG:06IT} instead of
\cite{Han&Kobayashi:81IT}.}. Generate $2^{nL_{12}}$ i.i.d
codewords $\nbf_1(\eta_1)$ for common messages according to
$\prod_{t=1}^np(n_1^{(t)}|q^{(t)})$ and randomly place them into
$2^{nR_{12}}$ bins\footnote{Random binning is used both for the
superposition block Markov encoding (relay) part and for DPC. To
distinguish, we use "cell" when referring to superposition block
Markov encoding and "bin" when referring to DPC.}; generate
$2^{nL_{10}}$ i.i.d codewords $\hbf_1(\omega_1)$ for cell indices
according to $\prod_{t=1}^np(h_1^{(t)}|q^{(t)})$ and randomly
place them into $2^{nR_{10}}$ bins; generate $2^{nL_{13}}$ i.i.d
codewords $\gbf_1(\psi_1)$ for cooperative messages according to
$\prod_{t=1}^np(g_1^{(t)}|q^{(t)})$ and randomly place them into
$2^{nR_{13}}$ bins. For each pair of
$(\nbf_1(\eta_1),\hbf_1(\omega_1))$, generate $2^{nL_{11}}$ i.i.d
codewords $\mbf_1(\xi_1,\eta_1,\omega_1)$ for private messages
according to $\prod_{t=1}^np(m_1^{(t)}|n_1^{(t)}h_1^{(t)}q^{(t)})$
and randomly place them into $2^{nR_{11}}$ bins.

To apply superposition block Markov encoding, we also need two
random partitions. Randomly place the above generated
$2^{nR_{23}}$ codewords $\wbf_2(k_2)$ into $2^{nR_{20}}$ cells,
and those $2^{nL_{13}}$ codewords $\gbf_1(\psi_1)$ into
$2^{nR_{10}}$ cells.

\emph{Encoding}: In block $b$, user 2 wants to send new indices
$i_{2b}, j_{2b}$ and $k_{2b}$. For cooperatively resolving the
remaining $Y_2$ uncertainty about $\wbf_{2}(k_{2,b-1})$ in the
previous block $b-1$, it also sends the cell index of
$\wbf_{2}(k_{2,b-1})$, denoted by $l_{2b}$. At the same time, user
1 wants to send new indices $i_{1b}, j_{1b}$, $k_{1b}$ and the
cell index of $\gbf_{1}(\psi_{1,b-1})$, denoted by $l_{1b}$. Since
user 1 can also perfectly calculate $l_{2b}$ at the end of block
$b-1$, it looks into bins $j_{1b}$, $k_{1b}$ and $l_{1b}$ for
codewords $\nbf_1(\eta_{1b})$ and $\gbf_1(\psi_{1b})$ and
$\hbf_1(\omega_{1b})$ that are jointly typical with
$\sbf_2(l_{2b})$, respectively. For the previously found
$(\nbf_1(\eta_{1b}),\hbf_1(\omega_{1b}))$, encoder 1 looks into
bin $i_{1b}$ for codeword $\mbf_1(\xi_{1b},\eta_{1b},\omega_{1b})$
such that
$(\qbf,\sbf_2(l_{2b}),\nbf_1(\eta_{1b}),\hbf_1(\omega_{1b}),\mbf_1(\xi_{1b},\eta_{1b},\omega_{1b}))$
are jointly typical.
%\bqa \nn
%\{\qbf,\sbf_2(l_{2b}),\nbf_1(j_{3b}),\hbf_1(l_{3b}),\mbf_1(i_{3b},j_{3b},l_{3b})\}
%\\\in
%A_{\epsilon}^{(n)}(QS_2N_1H_1M_1) \eqa where
%$A_{\epsilon}^{(n)}(\cdot)$ denotes the jointly typical set.
For the above bin searching, if there is more than one such
codeword, pick the one with the smallest index; if there is no
such codeword, declare an error. Then, user 1 sends $\xbf_1$
generated according to
$\prod_{t=1}^np(x_1^{(t)}|m_1(\xi_{1b},\eta_{1b},\omega_{1b})^{(t)}g_1(\psi_{1b})^{(t)}s_2(l_{2b})^{(t)}q^{(t)})$
and user 2 sends $\xbf_2$ generated according to
$\prod_{t=1}^np(x_2^{(t)}|v_2(i_{2b},j_{2b},l_{2b})^{(t)}w_2(k_{2b})^{(t)}h_1(\omega_{1b})^{(t)}q^{(t)})$.

\emph{Decoding}: User 2, as a relay to user 1, wants to correctly
recover the new index $k_{1b}$ sent in block $b$. Since it already
knows $\hbf_1(\omega_{1b})$ and $\sbf_2(l_{2b})$ during encoding,
it looks for all the sequences $\gbf_1(\psi_{1})$, such that \beq
\begin{array}{ll}
\{\qbf,\sbf_2(l_{2b}),\hbf_1(\omega_{1b}),\gbf_1(\psi_{1}),\tilde{\ybf}_{1b}\}\in
A_{\epsilon}^{(n)}(QS_2H_1G_1\tilde{Y}_1)\label{eq:E1}
\end{array}\eeq
If those $\gbf_1(\psi_{1})$ have the same bin index $k_{1b}$, we
declare $\hat{\hat{k}}_{1b}=k_{1b}$. Otherwise, we declare an
error. On the other hand, user 1 determines the unique
$\wbf_2(k_{2b})$, such that \beq
\begin{array}{ll}
\{\qbf,\sbf_2(l_{2b}),\hbf_1(\omega_{1b}),\wbf_2(k_{2b}),\tilde{\ybf}_{2b}\}\in
A_{\epsilon}^{(n)}(QS_2H_1W_2\tilde{Y}_2).
\end{array}\eeq
At the receiver side, we assume $Y_1$ knows $i_{1,b-1}$,
$j_{1,b-1}$, $l_{1,b-1}$ and $j_{2,b-1}$, and it can construct
$\mbf_1(\xi_{1,b-1},\eta_{1,b-1},\omega_{1,b-1})$,
$\nbf_1(\eta_{1,b-1})$, $\hbf_1(\omega_{1,b-1})$ and
$\ubf_2(j_{2,b-1})$, which are jointly typical with
$\ybf_{1,b-1}$. Now it wants to first decode bin indices $i_{1b},
j_{1b}, l_{1b}$ and the common message index $j_{2b}$. It looks
for $\mbf_1(\xi_1,\eta_1,\omega_1), \nbf_1(\eta_1),
\hbf_1(\omega_1), \ubf_2(j_2)$, such that \bqa\nn
\{\qbf,\mbf_1(\xi_1,\eta_1,\omega_1), \nbf_1(\eta_1),
\hbf_1(\omega_1), \ubf_2(j_2), \ybf_{1b}\}\\\in
A_{\epsilon}^{(n)}(QM_1N_1H_1U_2Y_1).\label{eq:E2} \eqa If those
sequences satisfying (\ref{eq:E2}) have the same bin indices and
message index respectively, we declare $\hat{i}_{1b}=i_{1b},
\hat{j}_{1b}=j_{1b}, \hat{l}_{1b}=l_{1b}$ and
$\hat{j}_{2b}=j_{2b}$. Otherwise, declare an error. Assuming cell
index $l_{1b}$ is successfully decoded at $Y_1$, then we declare
$\hat{k}_{1,b-1}=k_{1,b-1}$ if those sequences
$\gbf_1(\psi_{1})\in \Cmat(l_{1b})\cap \Dmat(y_1(b-1))$ have the
same bin index $k_{1,b-1}$. Here $\Cmat(l_{1b})$ denotes the set
of $\gbf_1(\psi_1)$ in cell $l_{1b}$, and $\Dmat(y_1(b-1))$ is the
ambiguity set, i.e., sequences of $\gbf_1(\psi_{1})$ such that
\bqa\nn \{\qbf,\mbf_1(\xi_{1,b-1},\eta_{1,b-1},\omega_{1,b-1}),
\nbf_1(\eta_{1,b-1}), \hbf_1(\omega_{1,b-1}), \\\ubf_2(j_{2,b-1}),
\gbf_1(\psi_{1}), \ybf_{1,b-1}\}\in
A_{\epsilon}^{(n)}(QM_1N_1H_1U_2G_1Y_1).\label{eq:E3} \eqa For
$Y_2$, the decoding process is the same and we skip the details.

\emph{Analysis of error probability}: We first consider
$P_{e,1}^{(n)}$ and we still use the story in block $b$. Let
$P_{0}$ denote the probability that there is no $\mbf$ in bin
$i_{1b}$, such that
$(\qbf,\sbf_2(l_{2b}),\nbf_1(\eta_{1b}),\hbf_1(\omega_{1b}),\mbf_1(\xi_{1},\eta_{1b},\omega_{1b}))$
are jointly typical. Then, \bqa P_0 &\leq&
(1-2^{-n(I(M_1;S_2|N_1H_1Q)+3\epsilon)})^{2^{n(L_{11}-R_{11})}}\\
&\leq& e^{-2^{-n(I(M_1;S_2|N_1H_1Q)+3\epsilon-L_{11}+R_{11}+1/n)}}
\eqa So, (\ref{eq:1}) guarantees $P_0\rightarrow 0$ as
$n\rightarrow \infty$. Similarly, bounds (\ref{eq:2})-(\ref{eq:4})
guarantee that encoder 1 can find codewords $\nbf_1(\eta_{1b})$,
$\gbf_1(\psi_{1b})$ and $\hbf_1(\omega_{1b})$, which are jointly
typical with $\sbf_2(l_{2b})$, respectively.

Now we calculate the error probability for user 2 (as a relay) to
decode $k_{1b}$. Denote the sent codeword $\gbf_1(\psi_{1b})$ as
$\gbf_1(k_{1b},k^*)$ since it is picked from bin $k_{1b}$. $k^*$
denotes the index of $\gbf_1(\psi_{1b})$ in bin $k_{1b}$. Let
$E_1(k_1,k)$ denote the event (\ref{eq:E1}) and let $P_1$ denote
the probability for user 2 to make a decoding error. Then \bqa
P_1&\equiv& Pr\{E_1^c(k_{1b},k^*)\hspace{.5cm} or \bigcup_{k_1\neq
k_{1b}}E_1(k_1,k)\}\\
&\leq& Pr\{E_1^c(k_{1b},k^*)\}+\sum_{k_1\neq
k_{1b},k}Pr\{E_1(k_1,k)\}\\
&\leq& \epsilon + \sum_{k_1\neq k_{1b},k}Pr\{E_1(k_1,k)\}
 \eqa
For $k_1\neq k_{1b}$, we know \bqn \lefteqn{Pr\{E_1(k_1,k)\}}\\
&=&\sum_{(\qbf\sbf_2\hbf_1\gbf_1\tilde{\ybf}_{1b})\in
A_{\epsilon}^{(n)}}p(\qbf)p(\gbf_1|\qbf)p(\sbf_2\hbf_1\tilde{\ybf}_{1b}|\qbf)\\
&\leq&
|A_{\epsilon}^{(n)}|2^{-n(H(Q)-\epsilon)}2^{-n(H(G_1|Q)-\epsilon)}2^{-n(H(S_2H_1\tilde{Y}_1|Q)-\epsilon)}\\
&\leq&
2^{-n(H(Q)+H(G_1|Q)+H(S_2H_1\tilde{Y}_1|Q)-H(QG_1S_2H_1\tilde{Y}_1)-4\epsilon)}\\
&\leq& 2^{-n(I(S_2H_1\tilde{Y}_1;G_1|Q)-4\epsilon)} \eqn
Therefore, $P_1\leq \epsilon +
2^{-n(I(S_2H_1\tilde{Y}_1;G_1|Q)-L_{13}-4\epsilon)}$. $\epsilon$
can be arbitrarily small by letting $n\rightarrow\infty$. Thus,
bound (\ref{eq:15}) assures $P_1\rightarrow 0$ as $n\rightarrow
\infty$.

For the decoding of $i_{1b}, j_{1b}, l_{1b}$ and $j_{2b}$ by
$Y_1$, it is a direct application of the simultaneous
superposition coding\cite{Han&Kobayashi:81IT}. However, regarding
our codebook generation scheme, particularly the construction of
$\mbf_1$, we will get a somewhat simpler description, similar to
that of Chong-Motani-Garg \cite{CMG:06IT}. This leads to the
bounds (\ref{eq:5})-(\ref{eq:13}) and we skip the details here.

Let $E_2(k_1,k)$ denote the event that $\gbf_1(k_1,k)\in
\Cmat(l_{1b})\cap \Dmat(y_1(b-1))$ and $E_{21}(k_1,k)$ denote the
event (\ref{eq:E3}). We also define an indicator function
$I(k_1,k)$. If $\gbf_1(k_1,k)$ satisfies (\ref{eq:E3}),
$I(k_1,k)=1$; otherwise, $I(k_1,k)=0$. The number of sequences in
$\Dmat(y_1(b-1))$ with bin index
$k_1\neq k_{1,b-1}$ is \bqa \lefteqn{||\Dmat(y_1(b-1)||}\\
&=&\sum_{k_1\neq k_{1,b-1},k}E(I(k_1,k))\\
&=&\sum_{k_1\neq k_{1,b-1},k}Pr\{E_{21}(k_1,k)\}\\
&\leq& \sum_{k_1\neq
k_{1,b-1},k}2^{-n(I(Y_1M_1N_1H_1U_2;G_1|Q)-4\epsilon)}\\
&\leq& 2^{-n(I(Y_1M_1N_1H_1U_2;G_1|Q)-L_{13}-4\epsilon)} \eqa Now,
let $P_2$ denote the probability of error for $Y_1$ to decode
$k_{1,b-1}$. Denote the actually sent codeword
$\gbf_1(\psi_{1,b-1})$ in block $b-1$ as $\gbf_1(k_{1,b-1},k^*)$.
Then, \bqa P_2&\equiv& Pr\{E_2^c(k_{1,b-1},k^*)\hspace{.5cm}or
\bigcup_{k_1\neq
k_{1,b-1}}E_2(k_1,k)\}\\
&\leq& \epsilon + \sum_{k_1\neq k_{1,b-1},k}Pr\{E_2(k_1,k)\}\\
&\leq& \epsilon+ ||\Dmat(y_1(b-1)||\cdot 2^{-nR_{10}}\\
&\leq& \epsilon+
2^{-n(I(Y_1M_1N_1H_1U_2;G_1|Q)+R_{10}-L_{13}-4\epsilon)}\eqa Bound
(\ref{eq:14}) guarantees $P_2\rightarrow 0$ as $n \rightarrow
\infty$. Thus, with bounds (\ref{eq:1})-(\ref{eq:15}), it is
guaranteed that $Y_1$ will correctly decode $i_{1b}, j_{1b},
l_{1b}, j_{2b}$ and $k_{1,b-1}$ at the end of block $b$ with a
probability arbitrarily close to 1. Then, the information state of
receiver $Y_1$ propagates forward, yielding the total decoding
error probability $P_{e,1}^{(n)}\rightarrow 0$ as $n\rightarrow
\infty$. The analysis of $P_{e,2}^{(n)}$ is similar to
$P_{e,1}^{(n)}$, which leads to bounds
(\ref{eq:16})-(\ref{eq:26}). Thus, $\Rmat_1^*$ is achievable.
Proof of $\Rmat_2^*$ is identical, hence $\Rmat^*$ is achievable
for ICC via time sharing.

The cardinality bound on $Q$, i.e., $|Q|\leq 33$ is obtained by
applying the Caratheodory Theorem to a set of inequalities that
bound the rate pair $(R_1,R_2)$, obtained via Fourier-Motzkin
elimination to Eqs. (\ref{eq:1})-(\ref{eq:26}).\hfill{Q.E.D.}\\

\emph{Remarks}: From the encoding-decoding strategy of the above
theorem, the achievable region $\Rmat^*$ is actually a
generalization of the HK region for IC, the capacity region of
degraded relay channels, and the capacity region of GVBC. It
reduces to those extreme cases under the conditions elaborated
below.

1)When the conferencing channel between the two users is very
poor, the bound in (\ref{eq:15}) and (\ref{eq:26}) can be very
small. In this case, allocating power for cooperation will
actually reduce the rates otherwise achievable via direct
transmission. As a result, the encoders will not allocate any
power to transmit $W_1$ and $W_2$, so $W_1=W_2=S_1=S_2=0$ and
$G_1=G_2=H_1=H_2=0$. Then, both $\Rmat_1^*$ and $\Rmat_2^*$ reduce
to the region in Proposition 1, which equals to the HK region.

2)When the conferencing channel between the two users is good
enough, it is not necessary to transmit messages directly to the
receiver, because cooperative transmission with the other user
will always yield a better rate. In this case, the encoders will
let $V_1=V_2=U_1=U_2=0$ and $M_1=M_2=N_1=N_2=0$. Now, if user 2
refrains from transmitting its own message and only serves as a
relay to user 1 (i.e., $W_2=S_2=G_2=H_2=0$), both $\Rmat_1^*$ and
$\Rmat_2^*$ reduce to the capacity region of the degraded relay
channel in Proposition 2. Similarly, if user 1 serves only as a
relay to user 2, it also reduces to the capacity region of the
degraded relay channel.

3)When the conferencing channel between the two users is ideal
(i.e., the conferencing channel capacity is infinite), the bounds
in (\ref{eq:15}) and (\ref{eq:26}) are no longer needed. So,
$W_1=W_2=G_1=G_2\approx 0$. Combining the result in case 2 that
$V_1=V_2=U_1=U_2=0$ and $M_1=M_2=N_1=N_2=0$, we can easily check
that $\Rmat_1^*$ reduces to \bqa R_1&\leq& I(Y_1;H_1|Q)-I(H_1;S_2|Q)\label{eq:case3:1}\\
R_2&\leq& I(Y_2;S_2|Q) \label{eq:case3:2}\eqa and $\Rmat_2^*$
reduces to \bqa
R_1&\leq& I(Y_1;S_1|Q)\label{eq:case3:3}\\
R_2&\leq& I(Y_2;H_2|Q)-I(H_2;S_1|Q)\label{eq:case3:4} \eqa For the
Gaussian case, the rate region
(\ref{eq:case3:1})-(\ref{eq:case3:2}) becomes (\ref{eq:Gaussian1})
and (\ref{eq:case3:3})-(\ref{eq:case3:4}) becomes
(\ref{eq:Gaussian2}). So, in this case, $\Rmat^*$ reduces to the
capacity region of GVBC.

4)During the review process of this paper, we became aware of
\cite{Tuninetti:07ITA}, which essentially tackles the same problem
using a different approach. In \cite{Tuninetti:07ITA}, user
cooperation results in a common information (in the sense of
\cite{Tan:80IC}) at the encoders and the decoder uses backward
decoding (similar to that of \cite{Sendonaris-etal:03COM}) instead
of the random partitioning (i.e., binning) we use in our approach.
Except for some extreme cases, it appears no subset relation can
be established. The obtained achievable region in
\cite{Tuninetti:07ITA} is simpler as it does not involve a large
number of auxiliary variables; however, the scheme in
\cite{Tuninetti:07ITA} is strictly suboptimal for certain extreme
cases (e.g., degraded relay channels, IC with degraded message
sets with weak interference, and MIMO BC) whereas our achievable
region can be easily shown to be optimal in each of these cases.

\section{Numerical Examples}
\begin{figure}[htp]
\centerline{\leavevmode \epsfxsize=2.5in \epsfysize=1.6in
\epsfbox{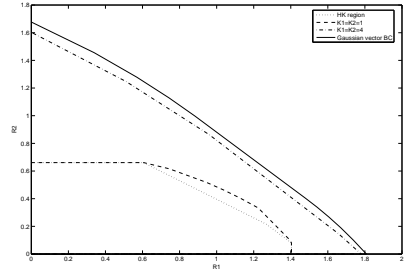}} \caption{\label{fig:result} Comparison of
$\Rmat$ with HK region and Gaussian vector broadcast channel
capacity. $P_1=6,P_2=1.5,a_{12}=a_{21}=0.74$}
\end{figure}

The standard form of a Gaussian interference channel is: \bqa Y_1&=&X_1+a_{21}X_2+Z_1\\
Y_2&=&a_{12}X_1+X_2+Z_2\eqa where $Z_1$ and $Z_2$ are arbitrarily
correlated zero mean, unit variance Gaussian random variables.
Suppose the power constraints of $X_1$ and $X_2$ are $P_1$ and
$P_2$, respectively. For the conferencing channel with perfect
echo cancellation, we have \bqa
\tilde{Y}_1=K_1X_1+\tilde{Z}_1,\hspace{1cm}
\tilde{Y}_2=K_2X_2+\tilde{Z}_2 \eqa where $\tilde{Z}_1$ and
$\tilde{Z}_2$ are both zero mean, unit variance Gaussian
variables. By reciprocity, we assume the channel coefficient
$K_1=K_2$. Since the computation of $\Rmat^*$ is formidable, here
we constrain all the inputs to be Gaussian distributed and set
$Q=\phi$ in order to compare our region with $\Gmat^{'}$ in (5.9)
of \cite{Han&Kobayashi:81IT} and the capacity region of GVBC. We
denote this modified region as $\Rmat$. Consider, that for certain
$\alpha_t, \beta_t, \gamma_t, \theta_t, \mu_t\in [0,1]$, with
$\alpha_t+\beta_t+\gamma_t+\theta_t+\mu_t=1$, where $t=1,2$, the
following hold: \bqa U_t\sim N(0,\beta_tP_t),W_t\sim
N(0,\gamma_tP_t), S_t\sim N(0,1)\\
V_t=V'_t+U_t+\sqrt{\theta_tP_t}S_t,
\hspace{0.2cm}\mbox{where}\hspace{0.2cm} V'_t\sim
N(0,\alpha_tP_t)\\
X_1=V_1+W_1+\sqrt{\mu_1P_1}S_2, X_2=V_2+W_2+\sqrt{\mu_2P_2}S_1
\eqa After applying Fourier-Motzkin Elimination on those bounds
(\ref{eq:1})-(\ref{eq:26}), we find that for each set of
$(\alpha_1, \beta_1, \gamma_1, \theta_1, \mu_1)$ and $(\alpha_2,
\beta_2, \gamma_2, \theta_2, \mu_2)$, both $S(Z_1)$ and $S(Z_2)$
are delimited by straight lines of slope $0, -\frac{1}{2}, -1, -2,
\infty$ as in the original HK region. Exhausting all the
parameters between $[0,1]$, and taking the convex hull of all
those $S(Z_1)$ and $S(Z_2)$, we get the achievable region $\Rmat$
for ICC in Fig.\ref{fig:result}.

\emph{Remarks}:

1) When there is no conferencing between the two users, the
achievable region reduces to the HK region. When the quality of
the conferencing channel improves, it increases our achievable
region for ICC within the limit of the capacity of GVBC.

2) When the channel coefficient is $K_1=K_2=4$, the region $\Rmat$
is already very close to the upper bound; when $K_1=K_2=1$, which
is equal to the channel coefficient of the transmitter to the
receiver, cooperation achieves a slightly better rate region than
independent transmission.

%3)There is no clear cut between the strategy of cooperative
%transmission and independent transmission. When the conferencing
%channel coefficient is within certain range, the boundary points
%of $\Rmat$ are achieved only when we combine those two strategies
%together, i.e., part of the messages are transmitted directly to
%the receiver while part of the messages are transmitted through
%cooperation.
3) For the channel coefficient $K_1=K_2=4$, the corresponding
relay channels (i.e., one of the users only serves as a relay) are
degraded, thus the intercepts of the bound at both axes are the
capacities of respective relay channels.

%\section{Conclusion}
%In this work, we investigate the problem of transmitter
%cooperation for interference channel (ICC), where each user can
%causally obtain part of the messages transmitted by the other user
%via conferencing. We find an achievable region for ICC, which
%reduces to the HK region, the capacity region of degraded relay
%channel, and the capacity of Gaussian vector broadcast channel
%under respective conditions.
%
%In the numerical example, we only computed $\Rmat$ by setting
%$Q=\phi$ and constraining the input to be Gaussian, which is a
%subset of our proposed region $\Rmat^*$ . We conjecture that, a
%larger rate region might be obtained if we allow $Q\neq\phi$.

\bibliographystyle{IEEEbib}
\bibliography{Journal,Conf,Misc,Book}

  \end{document}